\documentclass[final,5p,times,twocolumn,colorlinks=true,citecolor=blue,linkcolor=blue]{elsarticle}

\usepackage{graphicx}
\usepackage{dcolumn}
\usepackage{bm}
\usepackage{bbold}
\usepackage{amssymb,amsmath}
\usepackage{hyperref}

\usepackage{color}
\usepackage{amsfonts}
\usepackage{subfigure}
\usepackage{array}

\newcolumntype{L}{>{\centering\arraybackslash}m{3cm}}

\definecolor{bjcol}{rgb}{1,.44,0.13}

\definecolor{blue}{rgb}{0,0,1}

\definecolor{green}{rgb}{0,1,0}

\definecolor{red}{rgb}{1,0,0}

\definecolor{gray}{rgb}{.5,.5,.5}

\definecolor{darkgreen}{rgb}{.0,.5,.0}

\def\Fig#1{Fig.~\ref{#1}}

\def\Eq#1{Eq.~(\ref{#1})}

\def\eqref#1{(\ref{#1})}

\def\lA0{{\langle A_0 \rangle}}
\def\bA0{{\bar{A}_0}}

\def\0#1#2{\frac{#1}{#2}}

\graphicspath{{./figures/}{./}}

\begin{document}

\title{Chiral criticality and glue dynamics}

\author[ad:DUT]{Wei-jie Fu\corref{corauthor}}
\cortext[corauthor]{Corresponding author}
\ead{wjfu@dlut.edu.cn}

\address[ad:DUT]{School of Physics, Dalian University of Technology, Dalian, 116024,
  P.R. China}

\begin{abstract}

The chiral order-parameter $\sigma$ field and its higher-order cumulants of fluctuations are calculated within the functional renormalization group approach. The influence of the glue dynamics on the fluctuations of $\sigma$ field is investigated, and we find that the $\sigma$-field fluctuations are weakly affected by the glue dynamics. This is in sharp contrast to the baryon number fluctuations, which are sensitive to the glue dynamics and involve information of the color confinement. Implications of our calculated results on theoretical and experimental efforts to search for the QCD critical ending point are discussed.

\end{abstract}


\maketitle


Locating and searching for the QCD critical ending point (CEP) have attracted lots of attention in recent years. A nonmonotonic behavior of the kurtosis of the net proton multiplicity distribution with the change of the collision energy, has been observed in the beam energy scan (BES) program at the Relativistic Heavy-Ion Collider (RHIC) \cite{Adamczyk:2013dal,Luo:2015ewa,Luo:2017faz}, which might be attributed to critical fluctuations near the CEP. In order to pin down this possibility and, hopefully to resolve the existence and location of the CEP, reliable theoretical calculations and predictions are highly demanded.

Relevant theoretical studies put emphases on different sides. The first class is devoted to the computations and studies of equilibrium bulk properties of the quark-gluon plasma, in particular the fluctuations and correlations of conserved charges. Related studies involve, for instance, the first principle lattice QCD simulations \cite{Borsanyi:2013hza,Bellwied:2015lba,Ding:2015fca,Karsch:2017zzw,Bazavov:2017tot}, continuum functional approaches, e.g. the functional renormalization group (FRG) \cite{Fu:2015naa,Fu:2015amv,Fu:2016tey}, Dyson-Schwinger equations \cite{Qin:2010nq,Xin:2014ela,Gao:2016qkh}, etc. These studies are concerned with equilibrium critical fluctuations arising from the CEP. In heavy-ion collisions, besides the critical fluctuations, noncritical fluctuations also play a significant role \cite{Braun-Munzinger:2016yjz,Xu:2016skm,Li:2017via}, such as the participant or volume fluctuations for a given selection of the collision centrality, overall global conservation, acceptance cuts and so on. Furthermore, nonequilibrium critical fluctuations and their real-time evolution have attracted lots of attention in recent years \cite{Mukherjee:2015swa,Jiang:2015hri,Mukherjee:2016kyu,Jiang:2017sni}. It has been shown that higher-order critical fluctuations can be quite different from their equilibrium values \cite{Mukherjee:2015swa}, for more recent developments in this direction, see such as \cite{Stephanov:2017wlw,Stephanov:2017ghc}.

An promising approach to reveal the properties of the QCD chiral symmetry and its spontaneous dynamical breaking, is to study the chiral order parameter field, usually named as the $\sigma$-field, and its higher-order cumulants of fluctuations directly, see e.g. \cite{Stephanov:1999zu,Mukherjee:2015swa}. Besides the chiral symmetry and its breaking, however, the QCD is also characteristic of the color confinement, whose information is encoded in the glue dynamics. Furthermore, the net proton number fluctuations, which are employed to search for the QCD CEP in the BES program at RHIC, are physical quantities with the degree of freedom being baryons, rather than quarks. Therefore, it deserves to investigate the interrelations between the glue dynamics and the $\sigma$-field fluctuations. In this work, by employing the QCD-enhanced Polyakov--quark-meson (PQM) effective model \cite{Fu:2015naa}, we will study the influence of the glue dynamics on the $\sigma$-field fluctuations, and compare it with the net baryon number fluctuations. Computations in this paper are performed within the FRG approach, through which quantum and thermal fluctuations are encoded successively with the evolution of the renormalization group (RG) scale \cite{Wetterich:1992yh}. Interested readers are recommended to refer to, e.g., \cite{Pawlowski:2005xe,Pawlowski:2014aha,Pawlowski:2014zaa,Helmboldt:2014iya,Mitter:2014wpa,Braun:2014ata,Rennecke:2015eba,Fu:2015naa,Fu:2015amv,Wang:2015bky,Cyrol:2016tym,Fu:2016tey,Rennecke:2016tkm,Cyrol:2017ewj,Cyrol:2017qkl,Wang:2017vis,Fu:2017vvg} for more details about the FRG and recent QCD-related progresses.

 
Starting from an initial ultraviolet (UV) cutoff scale $\Lambda$, quantum, thermal and density fluctuations of different sizes, are successively encoded  within the FRG approach, with the RG scale $k$ running toward the infrared (IR) regime. Thus, a quantum theory is resolved with $k\rightarrow 0$. This process is described by the Wetterich equation \cite{Wetterich:1992yh} as follows 
\begin{align}
  \partial_{t}\Gamma_{k}[\Phi]&=\frac{1}{2}\mathrm{STr}\Big[\partial_{t} R_{k}\big(\Gamma_{k}^{(2)}[\Phi]+R_{k}\big)^{-1}\Big]\,,
 \label{eq:WetterichEq}
\end{align}
where $\Gamma_{k}$ is the scale-dependent effective action, $t=\ln (k/\Lambda)$; the super fields $\Phi$ consist of all field components of a theory, and the super trace runs over all degrees of freedom; $\Gamma_{k}^{(2)}$ is the second-order derivative of $\Gamma_{k}$ with respect to fields, i.e.,
\begin{align}
 \big (\Gamma_{k}^{(2)}[\Phi]\big)_{ij}:=\frac{\overrightarrow{\delta}}{\delta\Phi_i}\Gamma_{k}[\Phi]\frac{\overleftarrow{\delta}}{\delta\Phi_j}\,.
 \label{eq:Gamma2}
\end{align}
The regulator $R_{k}$ suppresses quantum fluctuations of size larger than $1/k$, while leaves others unchanged. 

%
\begin{figure*}[t]
\includegraphics[width=0.9\textwidth]{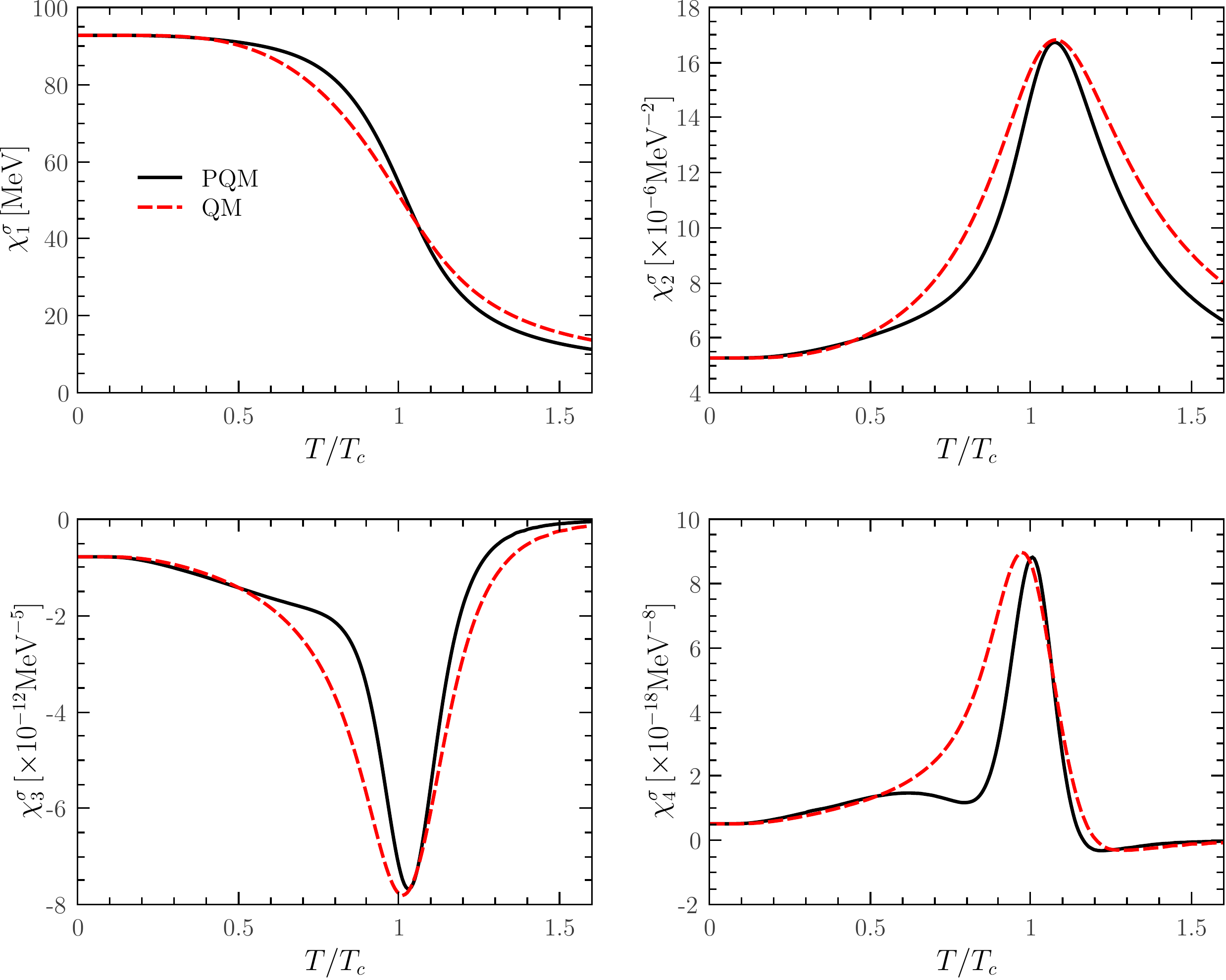}
\caption{Cumulants of the $\sigma$-field distributions, i.e., \Eq{eq:chis}, as functions of the temperature at vanishing chemical potential, calculated in the PQM and QM effective models.}\label{fig:chis}
\end{figure*}
%

Specifically as for the rebosonized QCD, the r.h.s of \Eq{eq:WetterichEq} is decomposed into the equation as follows
\begin{align}
  \partial_{t}\Gamma_{k}[\Phi]&=\frac{1}{2}\mathrm{Tr}\big(G^{AA}_{k}[\Phi]\partial_{t} R^{A}_{k}\big)-\mathrm{Tr}\big(G^{c\bar c}_{k}[\Phi]\partial_{t} R^{c}_{k}\big)\nonumber \\[2ex]
 &-\mathrm{Tr}\big(G^{q\bar q}_{k}[\Phi]\partial_{t} R^{q}_{k}\big)+\frac{1}{2}\mathrm{Tr}\big(G^{\phi\phi}_{k}[\Phi]\partial_{t} R^{\phi}_{k}\big)\,,
 \label{eq:WetterichEqQCD}
\end{align}
where $\Phi=(A, c, \bar c, q,\bar q, \phi)$, and the four terms on the r.h.s. relate to the gluon, ghost, quark, and hadronic degrees of freedom, respectively. Here $G$'s denote their propagators. For more discussions about the rebosonized QCD within the FRG approach, see e.g. \cite{Pawlowski:2005xe,Pawlowski:2014aha,Mitter:2014wpa,Braun:2014ata,Rennecke:2015eba,Cyrol:2017ewj,Fu:2018a}. In this work, we refrain from solving the whole coupled set of equations, and interested readers are recommended to refer to Ref. \cite{Fu:2018a}, in which flow equations of the rebosonized QCD are investigated at finite temperature and nonvanishing chemical potential. Instead of evolving the flows from an UV cutoff scale $\Lambda$ at the perturbative regime, such as $\sim 10^2$ GeV, we choose the initial scale $\Lambda \sim 1$ GeV, where the glue part, including the gluon and ghost, decouples from the matter part because of the large mass gap of the gluon \cite{Mitter:2014wpa,Braun:2014ata,Cyrol:2017ewj,Fu:2018a}. Thus, when the scale is below $\sim 1$ GeV, quantum fluctuations resulting from the gluon interactions are suppressed remarkably, and it is legal to separate the effective action into the glue and the matter parts, viz.,
\begin{align}
  \Gamma_k[\Phi]=\Gamma_{\text{\tiny{glue}},k}[\Phi]+\Gamma_{\text{\tiny{matt}},k}[\Phi]\,.\label{eq:Gasplit}
\end{align}
where the glue part corresponds to the first two terms on the r.h.s. of \Eq{eq:WetterichEqQCD}, and the matter part consists of quarks and hadrons, respectively. In this work, we focus on mesons for the hadronic degrees of freedom.

We adopt the following formalism for the scale-dependent effective of the matter, i.e.,
\begin{align} 
  \Gamma_{\text{\tiny{matt}},k}&=\int_{x}\Big\{Z_{q,k}\bar{q}\big[\gamma_{\mu}\partial_{\mu}-\gamma_{0}(\mu+igA_0)\big]q +\frac{1}{2}Z_{\phi,k}(\partial_{\mu}\phi)^2\nonumber\\[2ex]
  &\quad+h_{k} \,\bar{q}\left( T^{0}\sigma+i\gamma_{5} \bm{T}\cdot\bm{\pi}\right) q+V_{k}(\rho)-c\sigma\Big\}\,,\label{eq:action}
\end{align}
with notation $\int_{x}=\int_0^{1/T}d x_0 \int d^3 x$. $Z_{q,k}$ and $Z_{\phi,k}$ are the wave function renormalizations for the quark and meson, respectively. $h_{k}$ is the Yukawa coupling of the scalar and pseudo-scalar channels. $\bm{T}$'s are the generators of $SU(N_{f})$ in the flavor space with $\mathrm{Tr}(T^{i}T^{j})=\frac{1}{2}\delta^{ij}$, complemented with $T^{0}=\frac{1}{\sqrt{2N_{f}}}\mathbb{1}_{N_{f}\times N_{f}}$, with flavor number $N_{f}=2$.  $\phi=(\sigma,\bm{\pi})$ is the meson field, and the effective potential $V_{k}(\rho)$ with $\rho=\phi^2/2$ is $O(4)$ invariant. The term linearly proportional to the sigma field, i.e., $-c\sigma$, breaks the chiral symmetry explicitly, whose effects will be investigated in detail in the following.

In \Eq{eq:action} we also couple the quark field with the background gluon field, whose temporal component is nonvanishing at finite temperature. It has been well known that this background gluon field, also called as the Polyakov loop in a slight transformation, implements the QCD confinement through the $Z(3)$ symmetry and is indispensable to describe the QCD thermodynamics. Recent years have seen lots of progress on the study of the Polyakov dynamics, see e.g., \cite{Fukushima:2017csk} and references therein. Specifically, incorporation of the Polyakov dynamics and low energy effective models, leads to widely used Polyakov-loop-extended chiral models, for example the PQM \cite{Schaefer:2007pw} and the Polyakov--Nambu--Jona-Lasinio model \cite{Fukushima:2003fw,Ratti:2005jh,Fu:2007xc}.

In this work we will not directly solve the flow equation for $\Gamma_{\text{\tiny{glue}},k}$ in \Eq{eq:Gasplit}; instead, the QCD-enhanced glue potential $V_{\text{\tiny{glue}}}$, which is a polynomial in the Polyakov loop and has been discussed in detail in \cite{Fu:2015naa}, is employed. Therefore, the thermodynamical potential density reads
\begin{align}
  \Omega=V_{\text{\tiny{glue}}}+\Gamma_{\text{\tiny{matt}},\,k=0}\,,\label{}
\end{align}
where $\Gamma_{\text{\tiny{matt}},\,k=0}$ is obtained through integrating its relevant flow equation beginning from the initial scale $\Lambda$. Here we will not go into the details, for more discussions, see \cite{Fu:2015naa} for instance.

%
\begin{figure}[t]
\includegraphics[width=0.45\textwidth]{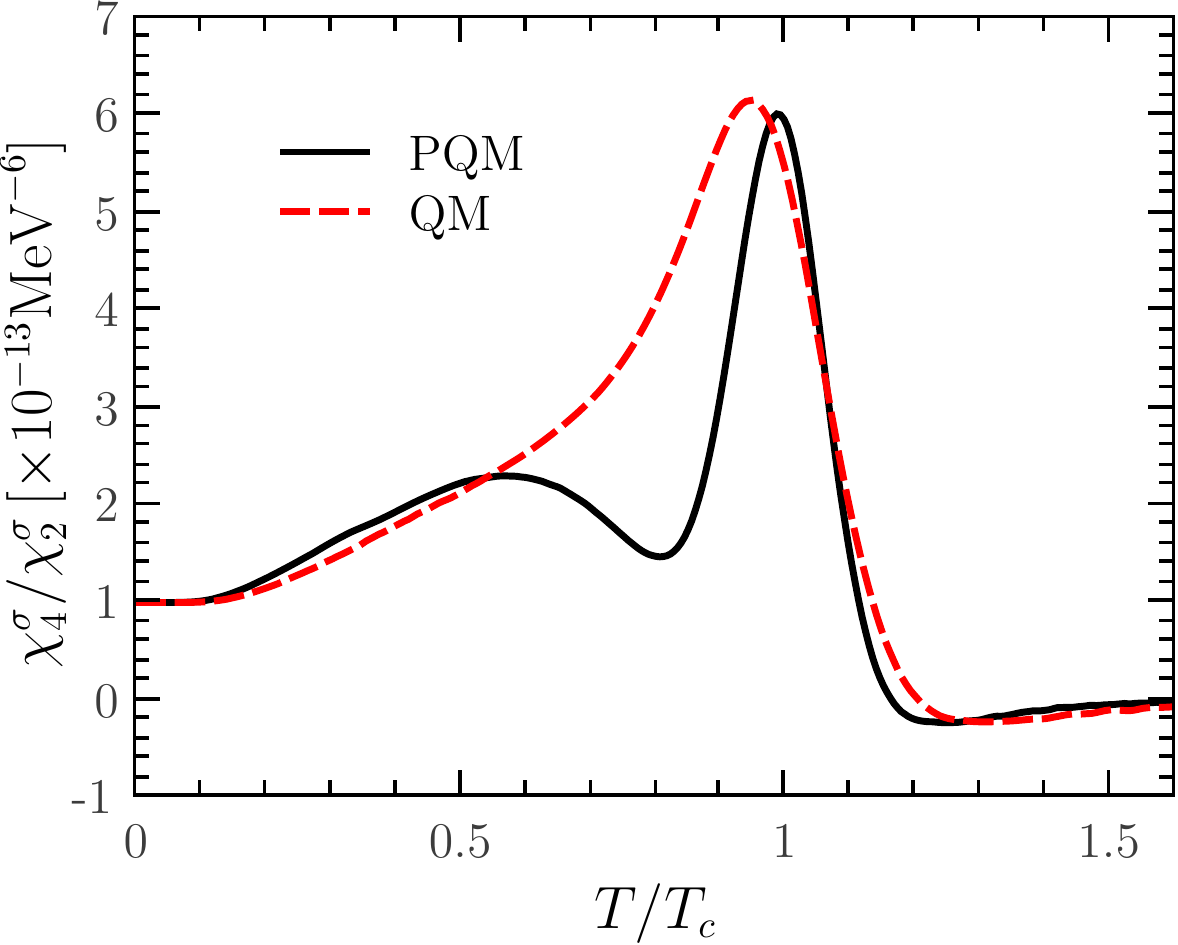}
\caption{$\chi_4^{\sigma}/\chi_2^{\sigma}$ as a function of the temperature calculated in the PQM and QM effective models.}\label{fig:R42s}
\end{figure}
%

In this work, we focus on two kinds of critical fluctuations, which are both important in the studies of the QCD phase transition and QCD phase structure. One is the fluctuations of the $\sigma$ field, which is the chiral order parameter and has been studied widely in literatures, see e.g., \cite{Stephanov:1999zu,Stephanov:2008qz,Mukherjee:2015swa}; the other is the fluctuations of the net baryon number or the net proton number, which is a significant probe to search for the QCD critical point in the experiments of BES, see \cite{Luo:2017faz} and references therein. More importantly, the emphasis is put on the influence of the glue dynamics on these two kinds of critical fluctuations.

The coefficient $c$ with nonvanishing value in the chiral symmetry breaking term in \Eq{eq:action}, on one hand, breaks the chiral symmetry explicitly and provides mass for the pions, on the other hand, works as an external source for the sigma field. Therefore, mean value and various cumulants of the sigma field could be obtained by differentiating the pressure $p=-\Omega$ with respect to the coefficient $c$, to wit
\begin{align}
  \frac{\partial p}{\partial c}&=\langle \sigma \rangle\,, \quad\quad  \frac{\partial^2 p}{\partial c^2}=\beta V\langle (\delta\sigma)^2 \rangle\,,\label{eq:chis12}\\[2ex]
  \frac{\partial^3 p}{\partial c^3}&=\Big(\beta V\Big)^2\langle (\delta\sigma)^3 \rangle\,,\\[2ex]
  \frac{\partial^4 p}{\partial c^4}&=\Big(\beta V\Big)^3\Big[\langle (\delta\sigma)^4 \rangle-3\langle (\delta\sigma)^2 \rangle^2\Big]\,,\label{}
\end{align}
with $\delta\sigma=\sigma-\langle \sigma \rangle$, the inverse temperature $\beta=1/T$ and the volume $V$, where the angle bracket denotes the ensemble average. For simplicity, we define
\begin{align}
  \chi_n^{\sigma}=\frac{\partial^n p}{\partial c^n}\,.\label{eq:chis}
\end{align}
For comparison, we also calculate the baryon number fluctuations which read
\begin{align}
  \chi_n^{B}=\frac{\partial^n }{\partial (\mu_B/T)^n}\frac{p}{T^4}\,.\label{eq:chiB}
\end{align}
with $\mu_B=3\mu$.

To precede the presentation of the calculated results, we would like to give a brief description about the numerical calculations. In this work we adopt the local potential approximation (LPA), i.e., $Z_{q,k}=Z_{\phi,k}=1$ and $\partial_t h_{k}=0$ in \Eq{eq:action}; for calculations beyond LPA, see, e.g., \cite{Fu:2015naa,Fu:2015amv,Fu:2016tey}. For the matter part, at the initial UV-cutoff scale $\Lambda$, which is chosen to being 700 MeV throughout this work, the effective potential in \Eq{eq:action} is symmetric and can be well approximated as follows
\begin{align}
  V_{\Lambda}(\rho)=\frac{\lambda_{\Lambda}}{2}\rho^2+\nu_{\Lambda}\rho\,.\label{}
\end{align}
with $\lambda_{\Lambda}=1$ and $\nu_{\Lambda}=(0.523\text{ GeV})^2$. These parameters, together with the Yukawa coupling $h=6.5$ and the coefficient of explicit chiral symmetry breaking $c=1.7\times 10^{-3}\,\text{GeV}^3$, yields hadronic observables at vacuum, which read the pion mass $m_{\pi}=135$ MeV, the sigma mass $m_{\sigma}=436$ MeV, the constituent quark mass $m_{q}=302$ MeV, and the $\pi$ decay constant which is identical to the vacuum expected value of the sigma field in \Eq{eq:chis12}, i.e., $f_{\pi}=\langle \sigma \rangle=92.8$ MeV. For the glue part, we employ the same QCD-enhanced glue potential as used in Refs. \cite{Fu:2015naa,Fu:2015amv}, which incorporates the back-reaction effect of the matte sector on the glue dynamics and thus leads to the correct scaling of the temperature, see \cite{Haas:2013qwp} for more details.

%
\begin{figure*}[t]
\includegraphics[width=0.9\textwidth]{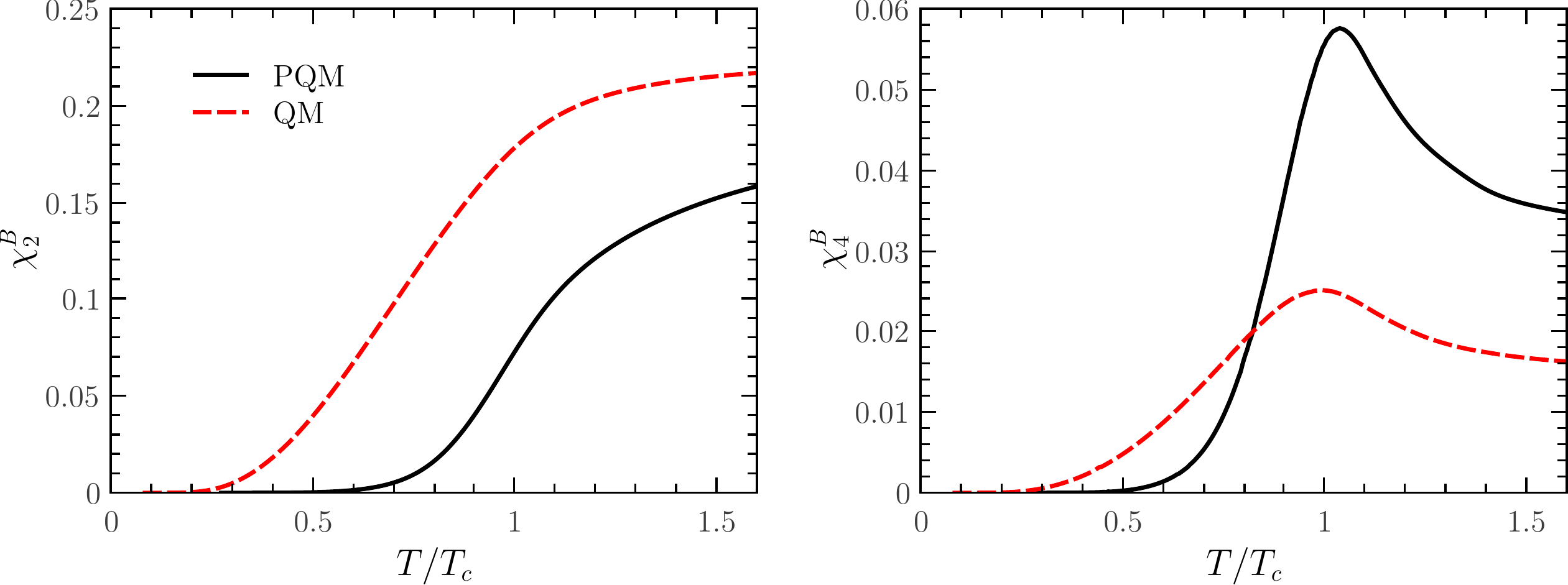}
\caption{Quadratic (left panel) and quartic (right panel) baryon number fluctuations defined in \Eq{eq:chiB} as functions of the temperature at vanishing baryon chemical potential, calculated in the PQM and QM effective models.}\label{fig:chiB}
\end{figure*}
%

%
\begin{figure}[t]
\includegraphics[width=0.45\textwidth]{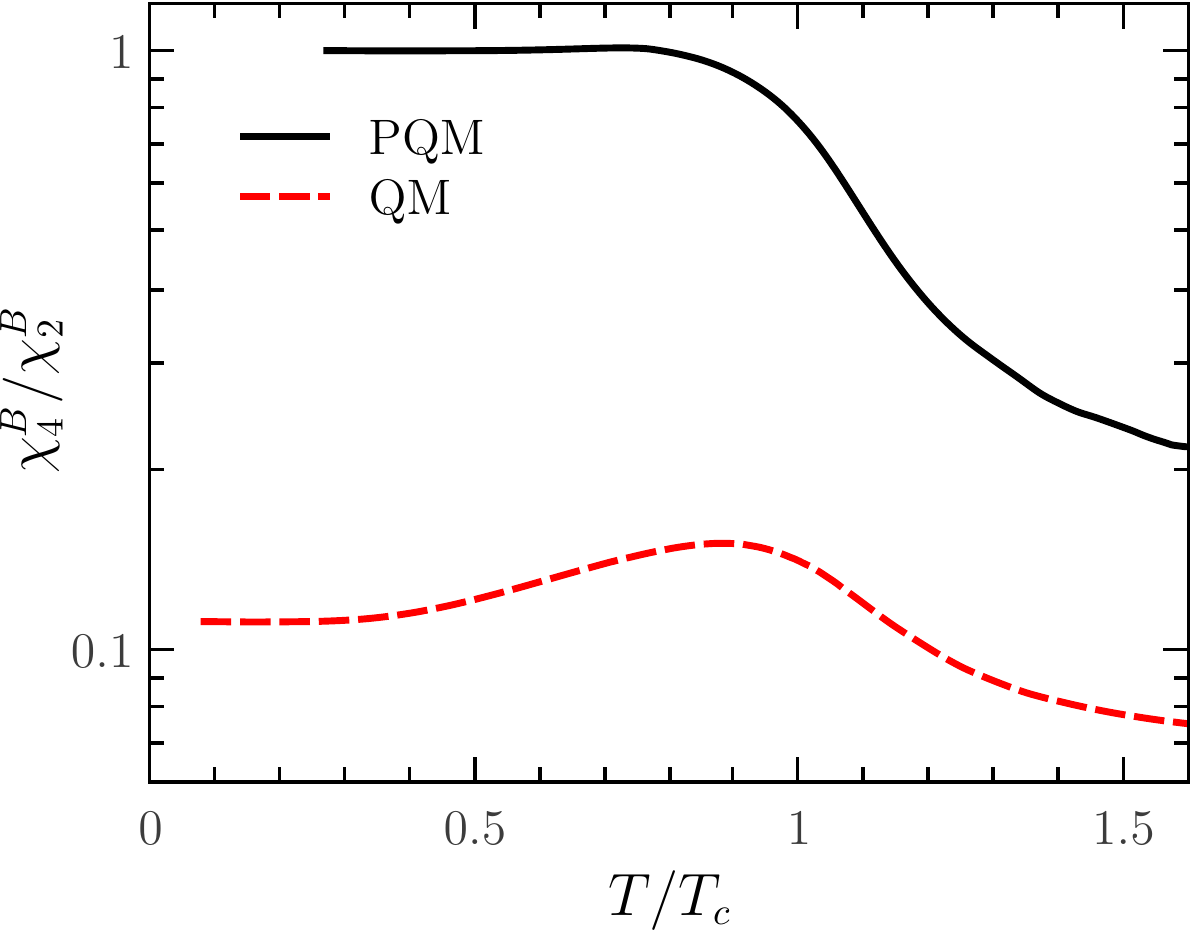}
\caption{$\chi_4^{B}/\chi_2^{B}$, i.e., the kurtosis of the net baryon number distribution, as a function of the temperature at vanishing baryon chemical potential, calculated in the PQM and QM effective models.}\label{fig:R42B}
\end{figure}
%

In \Fig{fig:chis} we show the first four orders of the fluctuations of the $\sigma$ field as defined in \Eq{eq:chis} as functions of $T$ in unit of $T_c$. The computations are performed in the PQM and quark-meson (QM) effective models, aimed at illuminating the role of the glue dynamics in the fluctuations of the $\sigma$ field, with the help of their difference. For the convenience of comparison, the temperature is rescaled by their pseudo-critical temperature $T_c$, which is determined by locating the maximal value of $\partial \langle \sigma \rangle/\partial T$ in this work. The value of $T_c$ is 183 MeV for the PQM and 154 MeV for the QM, respectively. Comparing $\langle \sigma \rangle$ calculated in the PQM and QM as shown in upper-left panel of \Fig{fig:chis}, one observes that the crossover regime shrinks when the glue dynamics is taken into account. This is easily understood and also expected in light of the physical implication of the Polyakov loop, which transforms the active degree of freedom at low temperature from quarks in the QM to baryons in the PQM. Therefore, higher temperature is needed to drive the occurrence of the chiral phase transition \cite{Fu:2007xc,Fu:2015naa}. In fact, this statement is even clearer when one sees higher-order fluctuations of the $\sigma$ field, for instance the quadratic, cubic, and quartic fluctuations of the $\sigma$ field, plotted in other panels of \Fig{fig:chis}. It is quite obvious that the crossover regime for the PQM is smaller than that in the QM. 

However, we should emphasize that the difference, discussed above about the $\sigma$-field fluctuations obtained in PQM and QM, is only quantitative and thus, is not of our concerns. On the contrary, the $\sigma$-field fluctuations calculated in these two effective models agree with each other qualitatively, if the unimportant quantitative difference is ignored. In another word, the $\sigma$-field fluctuations are almost not influenced by the glue dynamics, and the minor impact is indirect, such as through the modified pseudo-critical temperature by the glue potential. This situation is quite different from the baryon number fluctuations discussed in the following, which are closely related to, and affected directly by the glue dynamics. This is reasonable, since the $\sigma$ field is only related to the chiral property of the system, while does not encode the information of color confinement.

We show $\chi_4^{\sigma}/\chi_2^{\sigma}$ as a function of the temperature in \Fig{fig:R42s}. In the same way, we compare the result obtained in PQM and that in QM. It seems that the difference of the ratio between these two effective models is a bit larger than those in \Fig{fig:chis}, and there are two peaks on the curve for the PQM. Note, however, that only the second peak on the curve of PQM is relevant to the critical behavior, and the first wide bump is just due to the fact that the rapid change of the quartic $\sigma$-field fluctuation in PQM, as shown in the lower-right panel of \Fig{fig:chis}, is quite limited to a small region, thus leads to a stronger variance. Therefore, the first wide bump is not related to any critical behaviors. In a word, the qualitative behavior of the ratio $\chi_4^{\sigma}/\chi_2^{\sigma}$ is not affected by the glue dynamics as well.

In the following, we present some results about the baryon number fluctuations calculated in the LPA. These results are not state-of-the-art, and in fact there are lots of calculations beyond the approximation of LPA, see such as \cite{Fu:2015naa,Fu:2015amv,Fu:2016tey}. The purpose of computations of the baryon number fluctuations here is completely to facilitate the comparison between these two kinds of fluctuations. In \Fig{fig:chiB} we show the quadratic and quartic net baryon number fluctuations as functions of $T$ obtained in the PQM and QM effective models. Unlike the $\sigma$-field fluctuations, the baryon number fluctuations calculated in the PQM are remarkably different from those in the QM. In fact, if one sees the kurtosis of the baryon number distribution, i.e., $\chi_4^{B}/\chi_2^{B}$ in \Fig{fig:R42B}, it is immediately noticed that this difference is not quantitative, but rather qualitative. The baryon number fluctuation is not only sensitive to the chiral property, which is similar with the $\sigma$-field fluctuation as for this point, but also it is significantly influenced by the glue dynamics. It has been well known that the value of the ratio $\chi_4^{B}/\chi_2^{B}$ represents the degree of freedom of the system at low temperature, viz., $\chi_4^{B}/\chi_2^{B} \rightarrow 1$ is relevant to the baryons while $\chi_4^{B}/\chi_2^{B} \rightarrow 1/9$ to quarks \cite{Fu:2007xc,Fu:2015naa}, which correspond exactly to results obtained in the PQM and QM effective models, respectively, as shown in \Fig{fig:R42B}. Therefore, the confinement information is encoded in the PQM model through the glue dynamics, while the QM model does not include any confinement. Note that the kurtosis of the net baryon number distribution calculated in the PQM effective model, agrees well with lattice calculations, for more detailed discussions and comparison see, e.g., \cite{Fu:2015naa,Fu:2015amv,Fu:2016tey}. Obviously, computations of the baryon number fluctuations performed in a theoretical approach only with the chiral symmetry and its dynamical breaking while without the glue dynamic, such as the QM effective model, are far from sufficient to account for the lattice results and the experimental measurement.

In summary, in this work we have calculated the $\sigma$-field fluctuations within the FRG approach in the PQM and QM effective models. Emphasis is put on the interrelation between the $\sigma$-field fluctuations and the glue dynamics. We have found that the $\sigma$-field fluctuations are weakly influenced by the glue dynamics, which contrasts remarkably  with the case of the net baryon number fluctuations, that are strongly dependent on the glue dynamics and involve the color confinement information.

What does our calculated results imply for current efforts to search for the QCD CEP? We would like to address it from both the theoretical and experimental sides. Nonmonotonic behavior of the kurtosis of the net proton number distribution varying with the collision energy, has been observed in BES at RHIC \cite{Luo:2017faz}. Usually it is believed that the difference between the kurtosis of the proton number and that of the baryon number can be neglected, thus we assume the kurtosis of the baryon number distribution is identical to that of the proton number distribution. In order to explain experimental observations and make reasonable predictions, a number of interesting theoretical scenarios and approaches have been proposed and investigated in detail, such as equilibrium critical fluctuations, equilibrium noncritical fluctuations, nonequilibrium evolution, etc. Whatever the theoretical approaches are, the important thing is that the confinement information should be embedded in the theory besides the chiral property. The $\sigma$ field and its fluctuations alone are not sufficient to account for the baryon number fluctuations. 

On the contrary, since the QCD CEP arises from the chiral symmetry and its spontaneous dynamical breaking, rather not from the confinement-deconfinement phase transition, it is also a good idea to employ other physical quantities to search for the CEP in the experiments, which are only sensitive to the chiral symmetry, not affected by the color confinement, for instance the electric charge fluctuations and some physical observables which are only connected to the chiral order parameter field.

\section*{Acknowledgements}
We thank Yu-xin Liu for valuable discussions. The work was supported by the National Natural Science Foundation of China under Contracts Nos. 11775041.

\bibliography{ref-lib}

\end{document}